arXiv:1508.07428v2 [q-fin.ST] 16 Dec 2015

# Time-dependent scaling patterns in high frequency financial data

Noemi Nava[a,1], T. Di Matteo[b], Tomaso Aste[a,1]

[a] *Department of Computer Science, University College London, Gower Street, London, WC1E 6BT, UK*
[b] *Department of Mathematics, King's College London, The Strand, London, WC2R 2LS, UK*

## Abstract

We measure the influence of different time-scales on the dynamics of financial market data. This is obtained by decomposing financial time series into simple oscillations associated with distinct time-scales. We propose two new time-varying measures: 1) an amplitude scaling exponent and 2) an entropy-like measure. We apply these measures to intraday, 30-second sampled prices of various stock indices. Our results reveal intraday trends where different time-horizons contribute with variable relative amplitudes over the course of the trading day. Our findings indicate that the time series we analysed have a non-stationary multifractal nature with predominantly persistent behaviour at the middle of the trading session and anti-persistent behaviour at the open and close. We demonstrate that these deviations are statistically significant and robust.

## 1. Introduction

Financial markets are complex and dynamic systems that generate non-stationary, non-linear and noisy time series. These time series do not consist of a single driving force, but of several components that are superimposed

*Email addresses:* `n.morales.11@ucl.ac.uk` (Noemi Nava), `tiziana.dimatteo@kcl.ac.uk` (T. Di Matteo), `t.aste@ucl.ac.uk` (Tomaso Aste)

onto each other in a hierarchical form. The oscillating components are generated by many participants with diverse interests and investment horizons that interact and create effects that seem to repeat themselves cyclically with different characteristic periods [1]. These effects can be retrieved and quantified by looking at the statistical properties of financial data at different time-scales.

A classical approach to model financial markets has been the efficient market hypothesis (EMH) [2], which states that financial markets are efficient if they reflect all available information, thus arbitrage conditions are quickly eliminated. According to this theory, stock prices are unpredictable. The weak form of the EMH admits a rapid price adjustment process [2]. However, in practice, prices tend not to adjust to new information so rapidly, taking a certain amount of time. During this time investors can take actions to exploit temporary profitable opportunities arising from new information [3, 4].

According to the fractal market hypothesis (FMH) [5], financial participants are heterogeneous and market stability exists if there are investors with different time-horizons that create liquidity. Market participants have investment horizons that vary from seconds to years (market makers, noise traders, hedge funds). They treat the arriving information differently and affect the price dynamics in various ways depending on their trading time-scales [6]. The existence of investors with different investment horizons induces a stable market, markets become unstable when one horizon becomes dominant since liquidity dries up.

The FMH predicts that critical events are connected to dominant investment horizons and is usually associated to self-similarity, fractality or multifractality. Self-similarity in financial data was first studied by Mandelbrot [7, 8], and has been found to be present across financial markets with complex properties that are significantly related to economic and financial characteristics of the markets [9, 10, 11, 12, 13]. Self-similarity is related to the occurrence of similar patterns at different time-scales. In this sense, probabilistic properties of self-similar processes remain invariant when the process is viewed at different time-scales [7, 14]. A stochastic process $X(t)$ is statistically self-similar, with scaling exponent $0 < H < 1$, if for any real $a > 0$ it follows the scaling law:

$$X(at) \overset{d}{=} a^H X(t) \qquad t \in \mathbb{R}, \tag{1}$$

where the equality ($\overset{d}{=}$) is in probability distribution [15].

An example of self-similar process is fractional Brownian motion (fBm), a Gaussian process with stationary increments characterized by a positive scaling exponent $0 < H < 1$ [16]. When $0 < H < 1/2$, the increments of fBm show negative autocorrelation. The case $1/2 < H < 1$ corresponds to fBm with increment process exhibiting long range dependence, i.e., the autocorrelation of the increment process is positive. When $H = \frac{1}{2}$, the fBm reduces to Brownian motion (BM), a process with independent increments [15].

It has been observed that each moment of the distribution of financial returns varies as a power law of the time horizon with a different $H$ exponent [14]. This phenomenon is called multiscaling and reflects the occurrence of different dynamics at different time-scales, it can be attributed to the heterogeneity of market participants. Time-dependent scaling behaviour has also been observed in financial time series [17, 18], the local variations of roughness can be described by allowing the $H$ exponent to vary with time [19].

Entropy measures have also been used to measure the complexity of financial times series [20, 21, 22]. A low value of entropy indicates the presence of more predictable patterns that are therefore associated with periods of financial inefficiency. Conversely, when the time series exhibit more irregular and less predictable patterns, the uncertainty level is higher and such periods are described by larger values of entropy.

In this paper, we propose two new time-dependent measures based on the scaling properties of financial time series. These novel measures are obtained via the Hilbert-Huang transform [23], a methodology that as a first step, decomposes the analysed time series into several oscillatory modes by means of empirical mode decomposition (EMD). Secondly, the Hilbert transform is applied to these oscillations to obtain time varying attributes. The time-dependent scaling properties of financial time series are associated with the relative weights of the amplitudes at characteristic frequencies.

The first measure that we introduce in this paper is a scaling exponent that quantifies the relative hierarchical variations of the amplitudes of the components with respect to their associated time-scales; the second is an entropic measure quantifying the dispersion of the amplitudes of the components.

The remaining of this paper is organized as follows. In Section 2, we introduce the methodology used to identify the oscillating components of the data, the Hilbert-Huang transform. The proposed scaling measure with some application to self-similar process is described in Section 3. In Section 4, we propose a comparative entropy-like measure. In Section 5, we apply the proposed measures to intraday financial data. Finally, in Section 6, we provide conclusions and future perspectives.

## 2. Hilbert-Huang Transform

The Hilbert-Huang transform (HHT) [23] is a technique used to analyse non-linear and non-stationary time series. It was originally designed to study water wave evolution, but it has proven to be a useful tool for other complex signals, including financial time series [24, 25, 26, 27, 28].

The HHT consists of two steps: namely, empirical mode decomposition (EMD) and Hilbert transform (HT). The EMD decomposes the time series into a set of narrow-band intrinsic mode functions (IMFs) and the Hilbert transformation of these IMFs provides local frequency and amplitude attributes.

The EMD is a fully adaptive decomposition that in contrast to Fourier and wavelet transforms, does not require any a priori basis system [29]. Furthermore, it can be used to analyse non-stationary time series, it assumes that any time series consists of super-imposed oscillations. The purpose of the method is to identify these oscillations with scales defined by the local maxima and minima of the data itself. Hence, given a time series $x(t)$, $t = 1, 2, ..., T$, the EMD process decomposes it into a finite number of IMFs denoted as $c_k(t)$, $k = 1, ..., n$ and a residue function, $r(t)$. The number of IMFs, $n$, is approximately of the order of $\log_2(T)$ [23]. The IMFs are components oscillating around zero which are obtained through a sifting process that makes use of local extrema to separate oscillations starting with the highest frequency. At the end of the sifting process, the time series $x(t)$ can be expressed as:

$$x(t) = \sum_{k=1}^{n} c_k(t) + r(t). \tag{2}$$

The residue function, $r(t)$, is the non-oscillating drift of the data. For more details about this decomposition refer to [23]. The EMD was proposed as a way to pre-process time series before applying the Hilbert transform. It generates components of the time series whose Hilbert transform can lead to physically meaningful definitions of instantaneous amplitude and frequency. The Hilbert transformation of each function $c_k$ is defined as:

$$\hat{c_k}(t) = \frac{1}{\pi} \int_{-\infty}^{\infty} \frac{c_k(\tau)}{t-\tau}\, d\tau, \tag{3}$$

where the integral has a singular point at $\tau = t$ and it is defined as a Cauchy principal value. Each function $c_k$ and its Hilbert transformation, $\hat{c_k}$, produce a complex function $z_k$, defined by $z_k(t) = c_k(t) + \hat{c_k}(t)$ with amplitude

$$a_k(t) = \sqrt{c_k(t)^2 + \hat{c_k}(t)^2}, \tag{4}$$

and phase

$$\theta_k(t) = \tan^{-1} \frac{\hat{c_k}(t)}{c_k(t)}. \tag{5}$$

The instantaneous frequency is defined as the derivative of the phase with respect to time

$$\omega_k(t) = \frac{d\theta_k(t)}{dt}. \tag{6}$$

The instantaneous amplitude results in a smooth function, an 'envelope', that takes the overall shape of the time series taking the maxima and minima but never crossing the time series itself. This amplitude function represents the combined oscillations of all frequencies involved in the component but removes all the frequency information.

## 3. Time-dependent Scaling Exponent

The proposed time-dependent scaling exponent, denoted as $H^*(t)$, is constructed by observing the way local amplitudes $a_k(t)$ (Equation 4) change with respect to the local periods $\tau_k(t) = \omega_k(t)^{-1}$ (Equation 6) for all $k = 1, 2, \ldots, n$.

We first applied the method to fBm and observed that the amplitude function obtained through the HHT follows a power-law behaviour with respect to the instantaneous period:

$$a_k(t)^2 \propto \tau_k(t)^{2H^*(t)}, \tag{7}$$

where the exponent $H^*(t)$ describes the local scaling properties of the IMF amplitudes, and is comparable in magnitude to the self-similar exponent $H$.

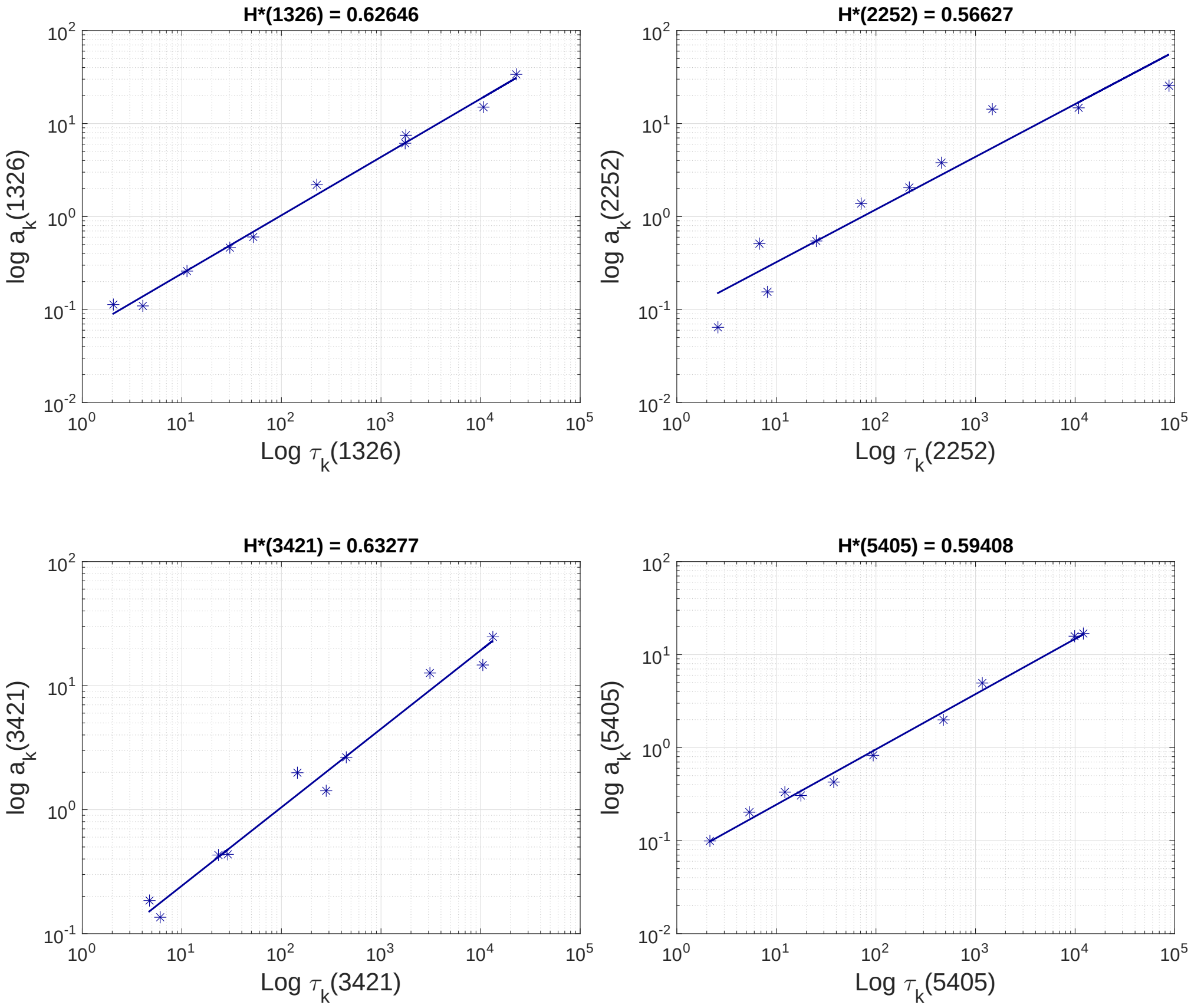

Figure 1: Illustration that for fBm the local amplitudes $a_k(t)$ and the periods $\tau_k(t)$ follow Equation 7: $a_k(t)^2 \propto \tau_k(t)^{2H^*(t)}$. Plots report instantaneous amplitude as a function of period for the following four randomly chosen times: $t = 1326, 2252, 3421, 5405$. The simulated process is a fBm with self-similar exponent $H = 0.6$ and length T=10,000 points. The straight lines are best-fit linear regressions.

In Figure 1, we report a particular instance of Equation 7 showing the linear fit between $\log a_k(t)$ and $\log \tau_k(t)$ for four randomly chosen times of a fBm with self-similar exponent $H = 0.6$ and length T=10,000. The values of $H^*(t)$ reported in the plots are obtained from the slope of the regression line. We observe that they are all consistently close to the self-similar value $H = 0.6$. For the chosen values of $t$, we calculated the goodness of the linear fit by estimating the coefficient of determination $R^2$ [30] (values of this coefficient range from 0 to 1, with 1 indicating a perfect fit between the data and the linear model). Results for the following four randomly

chosen times: $t = 1326, 2252, 3421, 5405$ are as follows: $R^2(1326) = 0.99$, $R^2(2252) = 0.90$, $R^2(3421) = 0.98$, $R^2(5405) = 0.99$, indicating therefore that for those instants of time the data are well represented by the log-linear model of Equation 7. Similar linear scaling results are obtained across all times, but the scaling exponent is different at each time step, making $H^*(t)$ a time-dependent estimator. A value of $H^*(t) > 0.5$ is obtained when around time $t$, the amplitude of long cycles is larger than in a pure random walk. This can be interpreted as a persistent behaviour in the amplitudes of the process, meaning that in a neighbourhood of time $t$ the process is in a cycle indistinguishable from a trend. On the contrary, values of $H^*(t) < 0.5$ represent a rougher and more chaotic behaviour around time $t$. These processes are composed of oscillations with more similar amplitudes across time-scales than in Brownian motion, creating a complex and uncertain behaviour. In this case, high frequency components are more active and their contribution to the total variance is more significant than in a random walk process.

### *3.1. Extended Simulation of Self-similar and Long Memory Processes*

To test the power-law relation of Equation 7, we extended the simulation set of fBm and we considered other two different self-similar processes, namely $\alpha$-stable Lévy motion (SLM) [15] and autoregressive fractionally integrated moving average (ARFIMA) processes [31].

For each stochastic process, we simulated $m = 1,000$ paths of length $T = 10,000$ points[1]. We estimated $H^*(t)$ and calculated the time-dependent sample mean over the number of simulations, i.e., we calculated $\langle H^*(t)\rangle = \frac{1}{m}\sum_{i=1}^{m} H_i^*(t)$.

We also estimated the sample mean of $H^*(t)$ over time and over the number of simulations, $\langle\langle H^*\rangle\rangle = \frac{1}{T}\sum_{t=1}^{T}\langle H^*(t)\rangle$. The standard deviation of $H^*(t)$ is calculated as:

$$std_{H^*} = \sqrt{\sum_{t=1}^{T}\sum_{i=1}^{m}\left(H_i^*(t) - \langle\langle H^*\rangle\rangle\right)^2 / (T \cdot m - 1)}.$$

For each process, the average over time and over the number of simulations of the coefficient of determination is also estimated and denoted as $\langle\langle R^2\rangle\rangle$. We compared the estimated $H^*(t)$ with the generalized Hurst exponent [33] with $q = 1$ and denoted as $H_G$.

[1]The length of the LFSM is set to $T = 2^{16} - 6000$, following the algorithm proposed in [32]

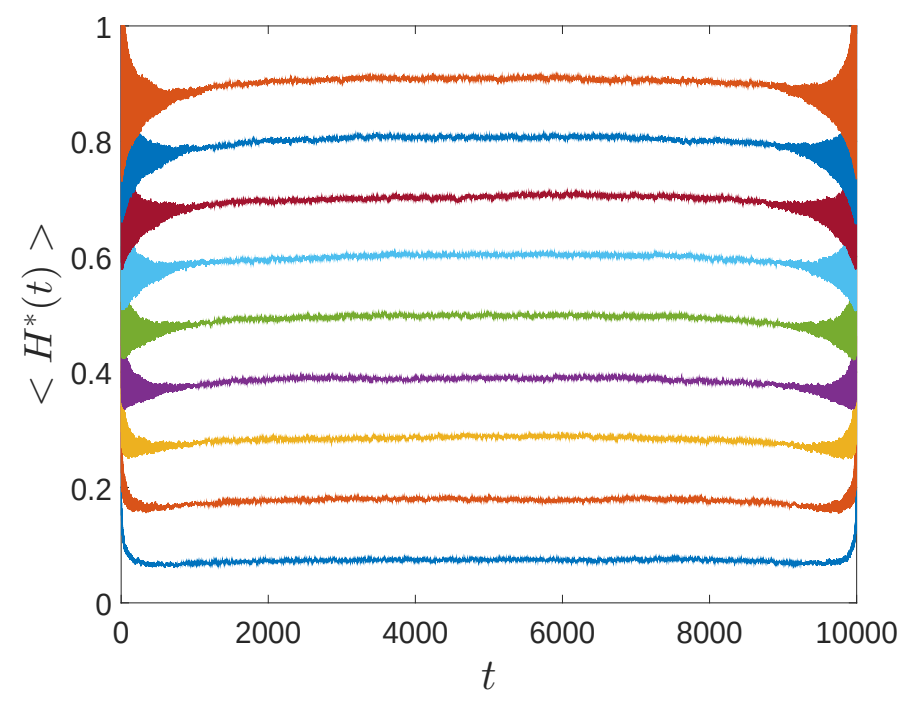

(a)

| **H** | $H^*$ | | | **GHE(1)** | |
|---|---|---|---|---|---|
| | $\langle\langle H^* \rangle\rangle$ | $std_{H^*}$ | $\langle\langle R^2 \rangle\rangle$ | $\langle H_G \rangle$ | $std_G$ |
| **0.1** | 0.07 | 0.08 | 0.20 | 0.15 | 0.01 |
| **0.2** | 0.18 | 0.08 | 0.39 | 0.22 | 0.01 |
| **0.3** | 0.28 | 0.08 | 0.59 | 0.31 | 0.01 |
| **0.4** | 0.38 | 0.09 | 0.73 | 0.40 | 0.01 |
| **0.5** | 0.50 | 0.09 | 0.81 | 0.50 | 0.01 |
| **0.6** | 0.60 | 0.09 | 0.86 | 0.60 | 0.01 |
| **0.7** | 0.70 | 0.10 | 0.89 | 0.70 | 0.01 |
| **0.8** | 0.80 | 0.10 | 0.92 | 0.79 | 0.01 |
| **0.9** | 0.90 | 0.11 | 0.93 | 0.87 | 0.01 |

(b)

Figure 2: a) Illustration that for fBm the scaling exponent $H^*(t)$ is on average close to the self-similar exponent $H$. The plot reports the sample mean of $H^*(t)$, denoted as $\langle H^*(t) \rangle$ and computed over $m = 1,000$ simulations of fBm with self-similar exponent $H = 0.1, 0.2, \ldots, 0.9$ (bottom to top) and length $T = 10,000$ points.
b) Sample mean of $H^*(t)$ over time and over number of simulations, denoted as $\langle\langle H^* \rangle\rangle$, standard deviation of $H^*(t)$ and sample mean of the coefficient of determination, $\langle\langle R^2 \rangle\rangle$. The values of GHE(1) denote the generalized Hurst exponent with $q = 1$.

- **Fractional Brownian motion**. Stochastic processes with scaling exponent varying from $H = 0.1, 0.2, ..., 0.9$ were simulated. All the simulations were done using the Matlab® wavelet toolbox. Results for different values of $H$ are reported in Figure 2(a). We observe that $\langle H^*(t) \rangle$ consistently varies around the input self-similar value of $H$. Some disturbances can be observed at the beginning and end of this time series due to the boundary effects of the EMD [34].

  In Table 2(b), we report $\langle\langle H^* \rangle\rangle$ and the standard deviation of $H^*(t)$. We observe a good agreement with the self-similar parameter $H$, but large values for $std_{H^*}$ that could be attributed to the local characteristics of $H^*(t)$. We notice that for fBm with scaling exponent $H < 0.3$, the $\langle\langle R^2 \rangle\rangle$ coefficients are small, indicating significant deviations from the scaling law of Equation 7. We obtained consistent results when comparing the estimated $H^*(t)$ with the generalized Hurst exponent.

  Moreover, we also considered fBm paths with shorter length, $T = 1,000$ and $T = 500$ points, we do not report these results, but we noted that the longer the time series, the better the estimation of the scaling exponent $H$. Likewise, the standard deviation and the goodness of the fit improve with the length of the time series. However, all results are consistent with the ones reported here for length $T = 10,000$.

- **$\alpha$-stable Lévy motion (SLM)**. This process is a generalization of the Brownian motion to the $\alpha$-stable distribution with $0 < \alpha \leq 2$. The case $\alpha = 2$ corresponds to Brownian motion. The SLM is a $1/\alpha$-self-similar process ($H = \frac{1}{\alpha}$ ) with stationary and independent increments [15]. Similarly, the extension of the fBm to the $\alpha$-stable distribution is the linear fractional stable motion (LFSM), a self-similar process that exhibits both heavy tails and serial correlation. The increments of LFSM have long range dependence when $H > 1/\alpha$ and negative dependence when $H < 1/\alpha$ [15]. Though the $\alpha$-stable Lévy motion is self-similar with $H \in [1/2, \infty)$, the self-similar parameter of LFSM varies in $(0, 1)$. For more details about these processes refer to [15].

  We generated SLM processes using the toolbox provided by [32], sample paths are of length $T = 10,384$, with parameters for the generation $m = 128$ and $M = 6000$, making $m(M + T)$ to be a power of 2, see [32] for more details. We considered the case $H = 1/\alpha$ for values of $H = 0.5, 0.55, \dots, 0.95$.

  The time-dependent sample mean over the number of simulations is displayed in Figure 3(a). We observe a noisier estimator than the one obtained for fBm. In Table 3(b), we report $\langle\langle H^*(t)\rangle\rangle$, noticing a fair approximation to the self-similar parameter $H$, with better results for processes with $H < 0.7$. The means of the coefficient of determination suggest that the scaling relation of Equation 7 is indeed satisfied. We compared the proposed estimator with the generalized Hurst exponent with $q = 1$, obtaining consistent results, i.e., $H_G = 1/\alpha$ [35].

- **ARFIMA**. We tested the log-linear relationship of Equation 7 in ARFIMA$(p, d, q)$ processes with Gaussian innovations, $p, q \in \mathbb{N}$, autoregressive and moving average coefficients respectively [31]. This model is an extension to ARIMA$(p, d, q)$ models, allowing the differencing exponent $d$ to take fractional values, $-1/2 < d < 1/2$. The correspondence between the two parameters $H$ and $d$ is given by $H = d + 1/2$. The interval $0 < d < 1/2$ of long range dependence corresponds to $1/2 < H < 1$ [36]. We considered the simple case of ARFIMA(0,d,0)

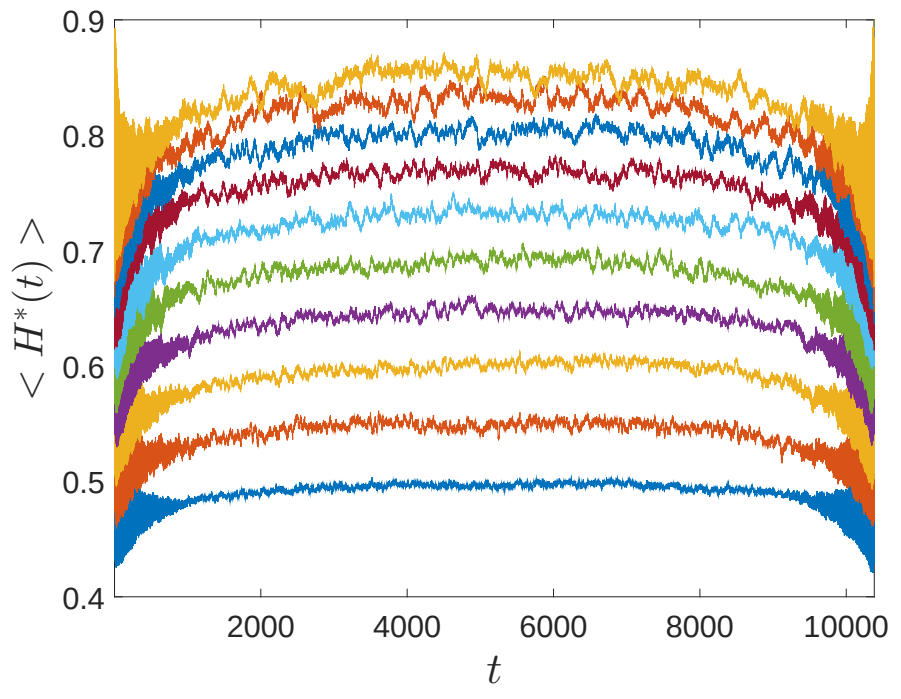

(a)

| **H** | $H^*$ | | | **GHE(1)** | |
|---|---|---|---|---|---|
| | $\langle\langle H^*\rangle\rangle$ | $std_{H^*}$ | $\langle\langle R^2\rangle\rangle$ | $\langle H_G\rangle$ | $std_G$ |
| **0.5** | 0.50 | 0.09 | 0.81 | 0.50 | 0.01 |
| **0.55** | 0.54 | 0.11 | 0.82 | 0.55 | 0.01 |
| **0.6** | 0.59 | 0.13 | 0.82 | 0.60 | 0.02 |
| **0.65** | 0.64 | 0.15 | 0.82 | 0.65 | 0.03 |
| **0.7** | 0.68 | 0.17 | 0.82 | 0.69 | 0.04 |
| **0.75** | 0.72 | 0.18 | 0.82 | 0.74 | 0.04 |
| **0.8** | 0.75 | 0.20 | 0.81 | 0.78 | 0.05 |
| **0.85** | 0.79 | 0.21 | 0.80 | 0.82 | 0.05 |
| **0.9** | 0.81 | 0.23 | 0.78 | 0.85 | 0.05 |
| **0.95** | 0.83 | 0.24 | 0.77 | 0.88 | 0.04 |

(b)

Figure 3: a) Illustration that for SLM the scaling exponent $H^*(t)$ is on average close to the value $H = \frac{1}{\alpha}$. The plots report the sample mean of $H^*(t)$, denoted as $\langle H^*(t)\rangle$ and computed over $m = 1,000$ simulations of SLM with self-similar exponent $H = 0.5, 0.55, \ldots, 0.95$ (bottom to top) and length $T = 10,000$ points.
b) Sample mean of $H^*(t)$ over time and over number of simulations, denoted as $\langle\langle H^*\rangle\rangle$, standard deviation of $H^*(t)$ and sample mean of the coefficient of determination, $\langle\langle R^2\rangle\rangle$.

with fractional order $d = -0.4, -0.3, \ldots, 0.4$ and length $T = 10,000$. We calculated $H^*(t)$ for the integrated ARFIMA time series.

From Figure 4(a), we observe that the estimator $\langle H^*\rangle$ is a good approximation of the exponent $H$. In Table 4(b), we report $\langle\langle H^*\rangle\rangle$, the standard deviation of $H^*(t)$ and the sample mean of the coefficient of determination, $\langle\langle R^2\rangle\rangle$. Similar to the fBm case, the estimation is more accurate for $H > 0.3$ and the coefficient of determination is closer to 1.

From the analysis of these three different processes, we observe that our proposed method produces a fair approximation to the self-similar parameter $H$. However, let us remark that this scaling exponent is not intended as an alternative method to estimate $H$, which can instead be obtained with more reliable tools [11, 37, 38]. The aim of this method is instead to compute the time-dependent amplitude contribution of the prevalent fluctuations present in a time series, distinguishing between periods when high or low frequencies are contributing more or less than what could be expected for Brownian motion.

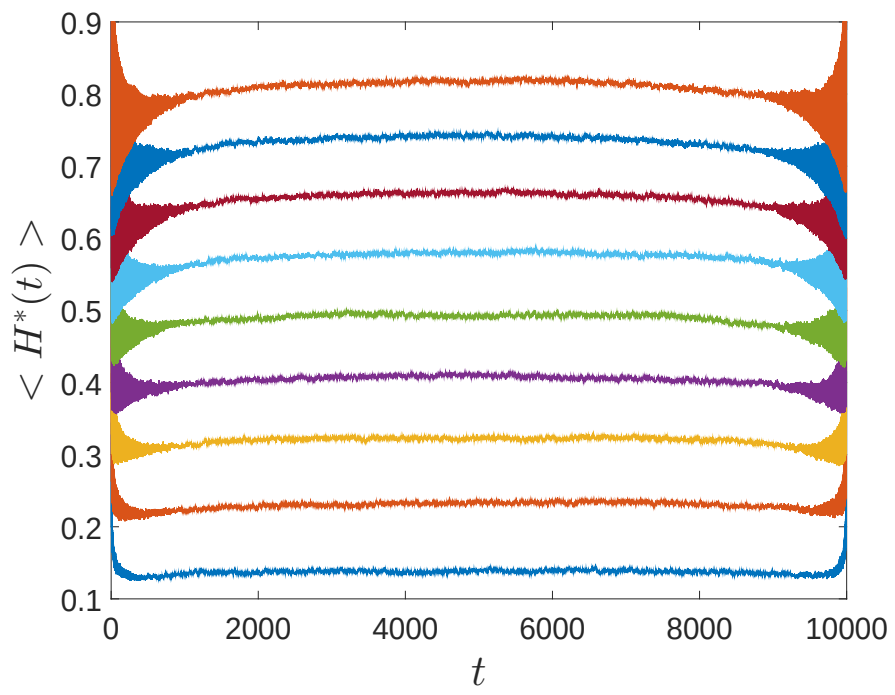

(a)

| **H** | $H^*$ | | | **GHE(1)** | |
|---|---|---|---|---|---|
| | $\langle\langle H^* \rangle\rangle$ | $std_{H^*}$ | $\langle\langle R^2 \rangle\rangle$ | $\langle H_G \rangle$ | $std_G$ |
| **0.1** | 0.14 | 0.09 | 0.29 | 0.21 | 0.01 |
| **0.2** | 0.23 | 0.09 | 0.48 | 0.27 | 0.01 |
| **0.3** | 0.32 | 0.09 | 0.64 | 0.34 | 0.01 |
| **0.4** | 0.40 | 0.09 | 0.74 | 0.42 | 0.01 |
| **0.5** | 0.49 | 0.09 | 0.81 | 0.50 | 0.01 |
| **0.6** | 0.57 | 0.09 | 0.85 | 0.59 | 0.01 |
| **0.7** | 0.65 | 0.09 | 0.88 | 0.68 | 0.01 |
| **0.8** | 0.73 | 0.10 | 0.90 | 0.77 | 0.01 |
| **0.9** | 0.81 | 0.10 | 0.91 | 0.84 | 0.01 |

(b)

Figure 4: a) Illustration that for ARFIMA(0,1,0) the scaling exponent $H^*(t)$ is on average close to the value $H = d + 0.5$. The plots report the sample mean of $H^*(t)$, denoted as $\langle H^*(t) \rangle$ and computed over $m = 1,000$ simulations of ARFIMA(0,d,0) with self-similar exponent $H = 0.1, 0.2, \ldots, 0.9$ (bottom to top) and length $T = 10,000$ points.
b) Sample mean of $H^*(t)$ over time and over number of simulations, denoted as $\langle\langle H^* \rangle\rangle$, standard deviation of $H^*(t)$ and sample mean of the coefficient of determination, $\langle\langle R^2 \rangle\rangle$.

## 4. Time-dependent Complexity Measure

We define a time-dependent Shannon entropy-like measure based on the square of the amplitude of the IMFs. This measure provides a time-varying quantification of complexity that offers an alternative to the scaling exponent to measure the strength of cycles present in financial time series. Making use of the functions $a_k$, described in Equation 4, we define a time-scale relative distribution of amplitudes as:

$$p_k(t) = \frac{a_k^2(t)}{\sum_{k=1}^{n} a_k^2(t)}, \tag{8}$$

where $n$ is the number of IMFs excluding the residue. Similarly to Shannon entropy [39], we define the time-dependent complexity measure as:

$$C^*(t) = -\sum_{k=1}^{n} p_k(t) \ln p_k(t). \tag{9}$$

Equation 9 provides a measure of the distribution of amplitudes between the oscillating components. If the total amplitude at time $t$ is concentrated in one oscillation mode, we observe a low complexity value, implying that around time $t$ the process is following a prevalent trend. On the contrary, if at time $t$ all the oscillation modes have similar amplitudes, we obtain a large complexity value that indicates a more erratic and unpredictable behaviour.

Thus, $C^*(t)$ provides a time-varying estimation of disorder and adapts closely to our visual perception of complexity. Moreover, Equation 9 offers a more general measure of uncertainty than the variance since the latter measures the dispersion around the mean, while $C^*(t)$ measures the dispersion of energy around the different IMFs. Similar to an entropy measure, the value of the proposed complexity at time $t$ varies between zero, if one IMFs dominates the energy of the process, and $\log(n)$ if the energy is uniformly distributed between the $n$ IMFs.

The choice of weights equal to the square of the amplitudes in Equation 8 is arbitrary, although it is in agreement with other measures of entropy that have been defined, for example in [40]. We tested alternative choices, such as the linear weight $a_k(t)$, obtaining analogous results to the ones reported here.

Let us note that, although not independent, the two measures convey different information. The estimator $C^*(t)$ is an information quantifier of uncertainty, it is obtained from the distribution of the amplitudes regardless of their time-scales and it only quantifies the homogeneity of the components. On the other hand, the scaling exponent, $H^*(t)$, measures the change in the amplitudes across time-scales, testing the scaling law in Equation 7. In this respect, $H^*(t)$ is a more restrictive measure that assumes a log-linear relationship between amplitudes and periods.

## 5. Time-dependent Scaling in Financial Markets

We applied the proposed measures to intraday prices of four stock indices: (1) S&P 500 (USA), (2) IPC (Mexico), (3) Nikkei 225 (Japan) and (4) XU 100 (Turkey). We intentionally chose two financial markets that are classified as developed (USA and Japan) and two emerging markets (Mexico and Turkey) with the additional feature that the Japanese and Turkish stock exchanges have two trading sessions separated by a lunch break.

The data set, obtained from Bloomberg, consists of prices recorded at 30-second intervals. It covers a period of 5 months, from January $15^{th}$, 2014 to June $16^{th}$, 2014. The number of days and the number of data points for every trading day depend on the opening hours of each stock exchange.

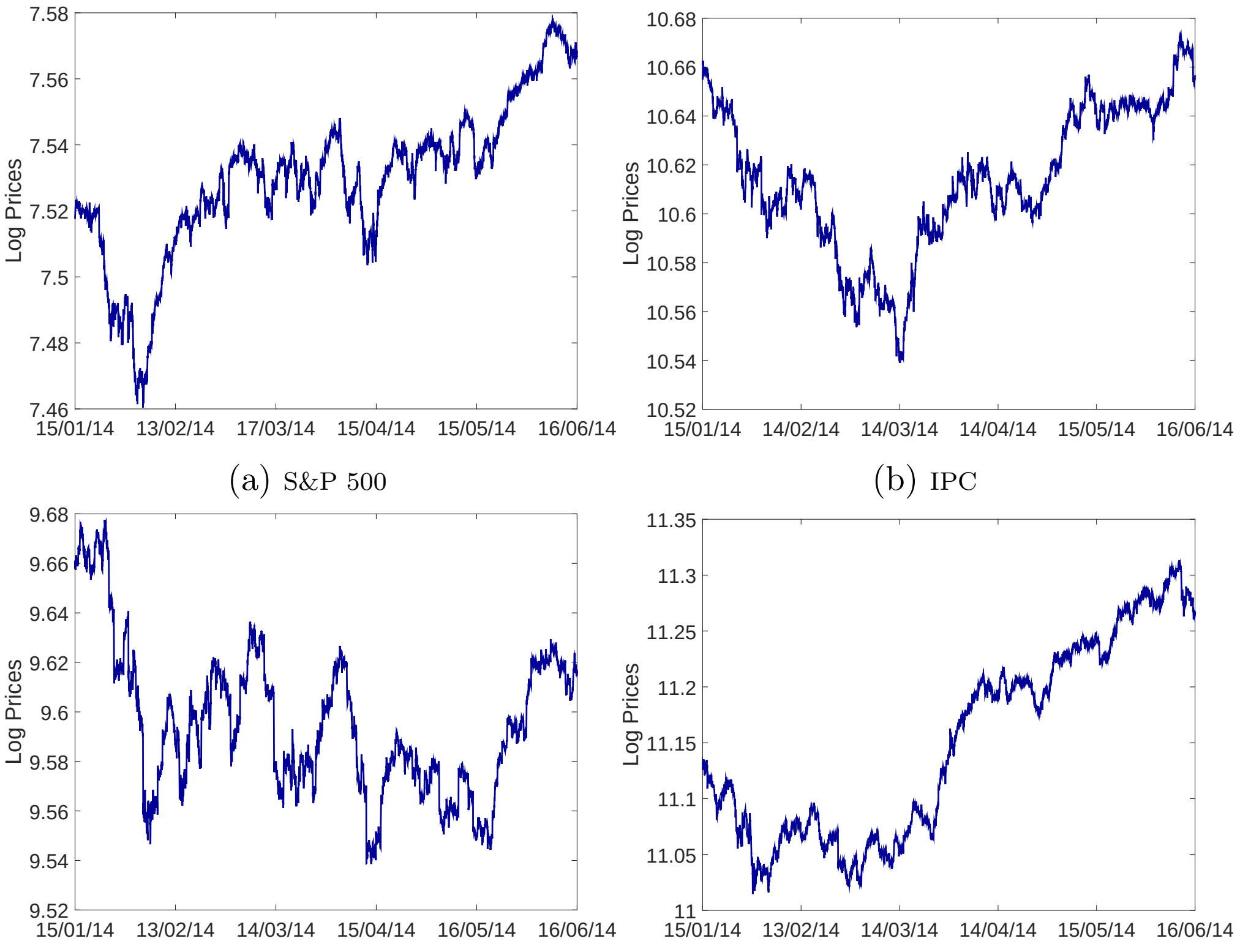

(a) S&P 500 (b) IPC

(c) Nikkei 225 (d) XU 100

Figure 5: 30-second sampled log-prices for different stock indices for the period January $15^{th}$, 2014 to June $16^{th}$, 2014. (a) S&P 500, (b) IPC, (c) Nikkei 225 and (d) XU 100.

Logarithmic prices of each financial stock index are plotted in Figure 5. Table 1 shows the number of days and the length of each time series.

| Country | Index | No. of Days | Length |
|---|---|---|---|
| USA | S&P 500 | 105 | 81,900 |
| Japan | Nikkei 225 | 104 | 62,400 |
| Mexico | IPC | 101 | 78,780 |
| Turkey | XU 100 | 106 | 78,440 |

Table 1: Number of days and length of each financial time series.

The time evolution of $H^*(t)$ over the 5-month period for the four financial indices is shown in Figure 6 (dark-blue line). We note that $H^*(t)$ has large intraday variations that make difficult to identify any trend over longer periods. For this reason, we report a moving average version of $H^*(t)$ denoted as $\bar{H}^*(t)$ (light-blue line in the same Figure). More specifically, $\bar{H}^*(t)$ is calculated from the relation, $\bar{a}_k(t)^2 \propto \bar{\tau}_k(t)^{2\bar{H}^*(t)}$, where $\bar{a}_k(t)$ and $\bar{\tau}_k(t)$ are the averages over a rolling window of the size of a trading day. The dashed red line in this Figure indicates the value $H = 0.5$.

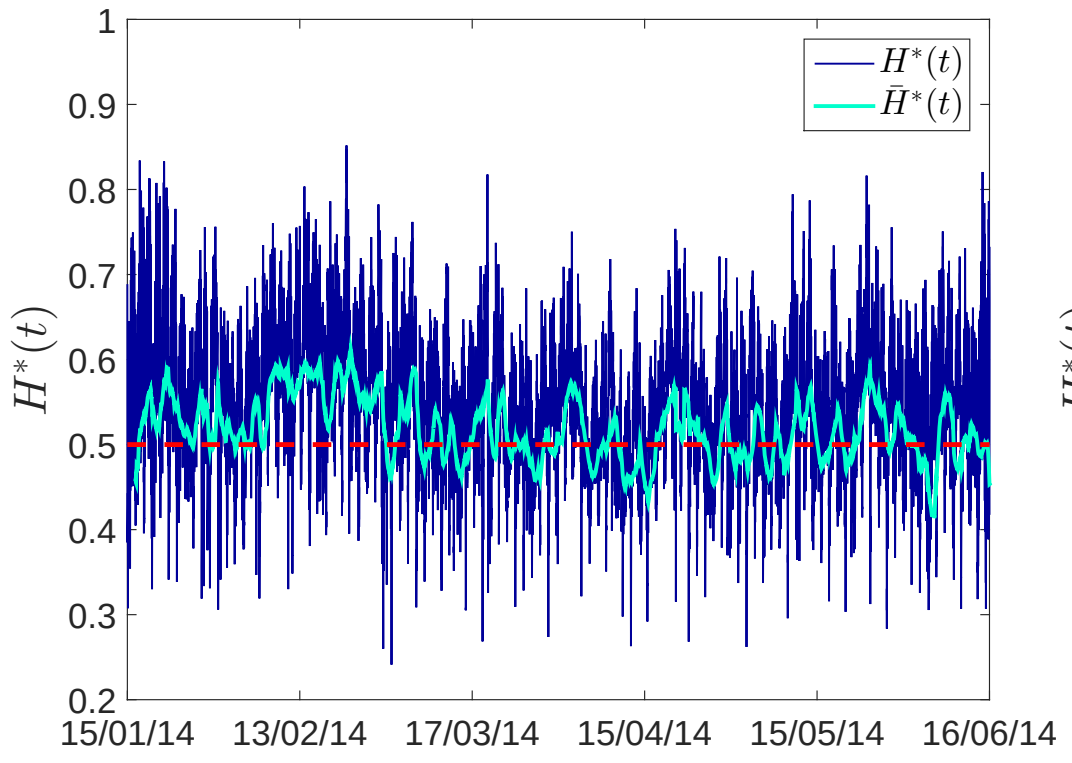

(a) S&P 500

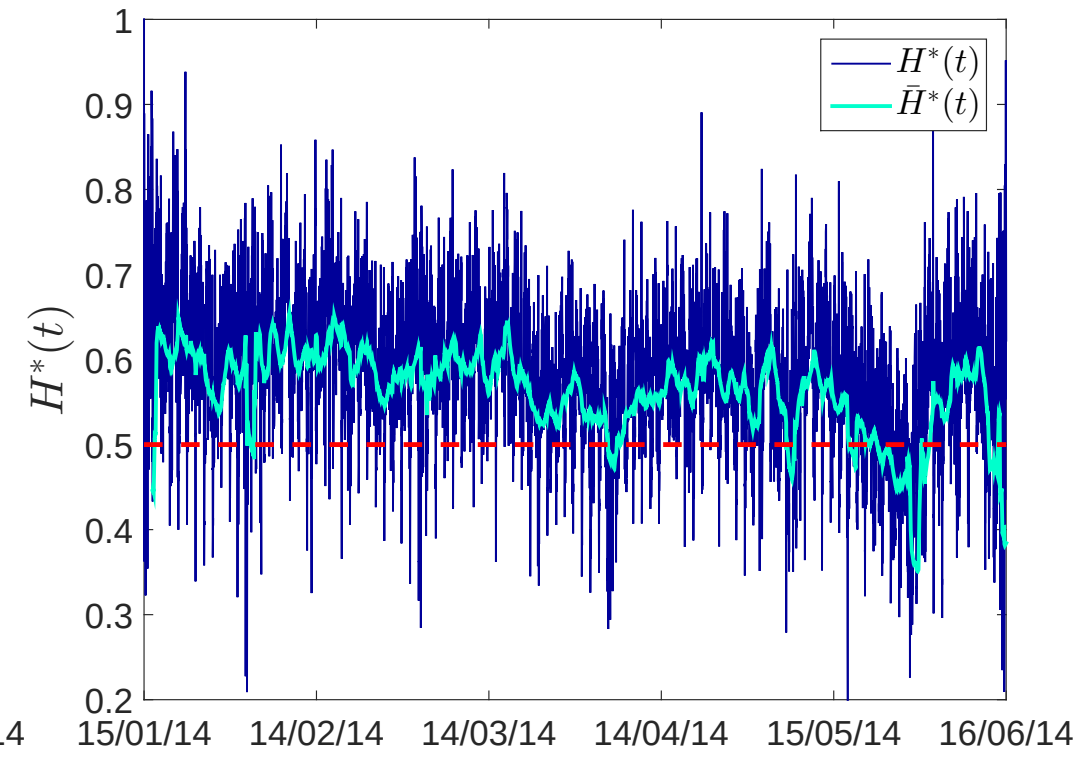

(b) IPC

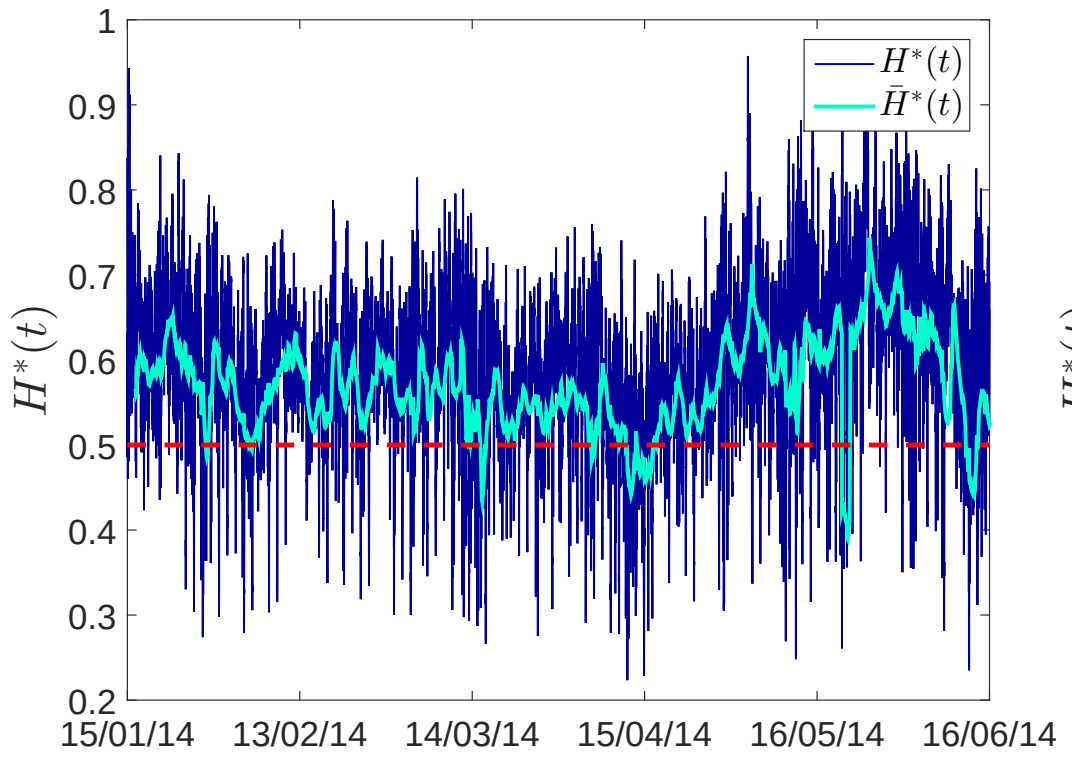

(c) Nikkei 225

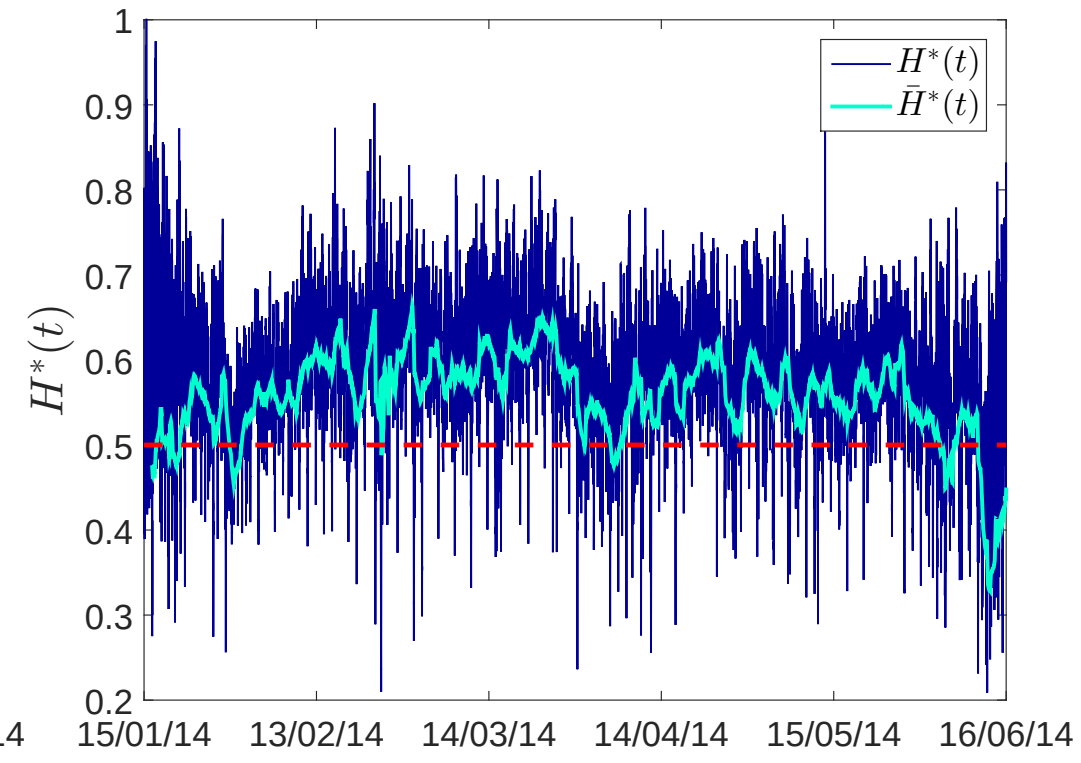

(d) XU 100

Figure 6: Time-dependent scaling exponent for different stock indices for the period January $15^{th}$, 2014 to June $16^{th}$, 2014. The scaling exponent $H^*(t)$ is depicted by a dark-blue line. The light-blue line represents $\bar{H}^*(t)$, a rolling average over the length of a trading day. The red line indicates the value $H = 0.5$. (a) S&P 500, (b) IPC, (c) Nikkei 225 and (d) XU 100.

By comparing the values of $H^*(t)$ and $\bar{H}^*(t)$ for the four different stock indices, we observe that the S&P 500 index is the one closest to the value $H = 0.5$, that assuming a Gaussian distribution will imply BM behaviour, see Figure 6(a). In this Figure, $\bar{H}^*(t)$ fluctuates around $H = 0.5$ exposing only some brief departures from it. For instance, we can detect a period around February 2014 where the scaling parameter results in significantly larger values. In this period of time, the S&P 500 index was indeed in a rising rally, see Figure 5(a). Therefore, this suggests that the identified persistent behaviour could be attributed to the recognition of a long time-scale cycle with larger amplitudes than in the case of pure random walk.

In Figure 6(c), we report the scaling dynamics for the Nikkei 225 index. We observe that $\bar{H}^*(t)$ has values constantly above 0.5, specially at the end of the analysed period. It should be noted that this market has lunch breaks that affect the intraday values of $\bar{H}^*(t)$.

For the IPC index the values of $H^*(t)$ and $\bar{H}^*(t)$ are consistently closer to $H = 0.6$ than to the BM value $H = 0.5$, see Figure 6(b). This suggests that IPC returns show intervals where the amplitude displays a persistent behaviour. Similarly, the Turkish scaling exponents take values larger than $H = 0.5$, see Figure 6(d).

We tested the validity of Equation 7 when applied to financial data by computing for every time $t$ the coefficient of determination $R^2(t)$. The mean over the whole period is reported in the second column of Table 2. We also considered the three cases: $H^*(t) < 0.45$, a window around $0.45 < H^*(t) < 0.55$ and $H^*(t) > 0.55$. We observe that the goodness-of-fit is generally better for $H^*(t) > 0.5$, the interesting case when financial data show trending behaviour, see Table 2.

| Index | $\langle R^2 \rangle_{All}$ | $\langle R^2 \rangle_{H^*<0.45}$ | $\langle R^2 \rangle_{0.45<H^*<0.55}$ | $\langle R^2 \rangle_{H^*>0.55}$ |
|---|---|---|---|---|
| S&P 500 | 0.8753 | 0.825 | 0.8716 | 0.8916 |
| IPC | 0.8812 | 0.7971 | 0.8703 | 0.8915 |
| NIKKEI 225 | 0.8072 | 0.7345 | 0.7829 | 0.8198 |
| XU100 | 0.9196 | 0.7987 | 0.8737 | 0.9209 |

Table 2: Average goodness-of-fit coefficient ($R^2$) for the amplitude versus period log-linear model for different financial indices. First, the average is calculated for all the times $t$, then it is calculated separately for those times where $H^*(t) < 0.45$, $0.45 < H^*(t) < 0.55$ and $H^*(t) > 0.55$.

For a comparative analysis, we calculated the complexity measure $C^*(t)$ described by Equation 9. The obtained values for each financial stock index are illustrated in Figure 7. The complexity values for S&P 500 index, Figure 7(a), are overall the largest among the four indices.

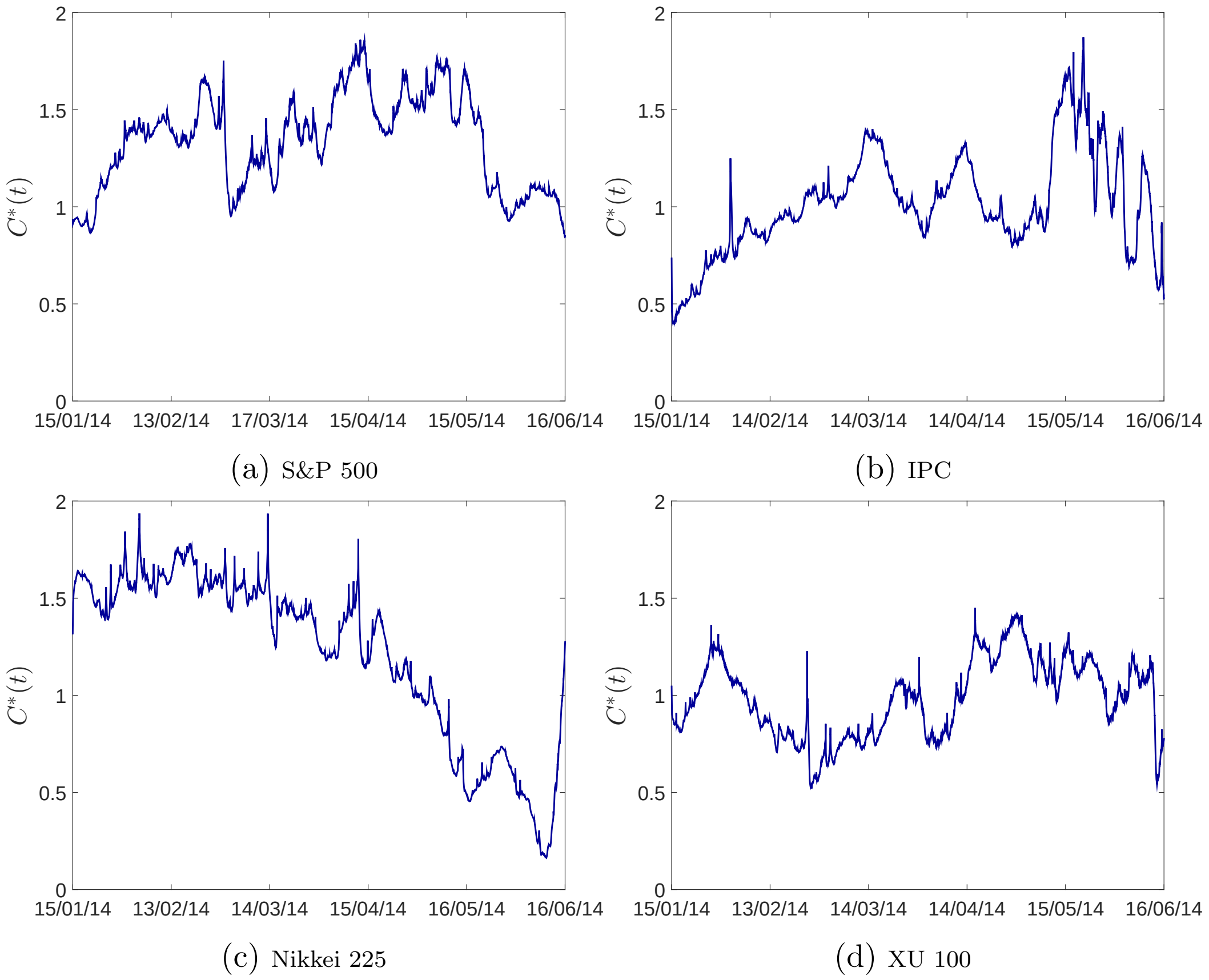

Figure 7: Time-dependent complexity measure, $C^*(t)$, for different stock market indices for the period January $15^{th}$, 2014 to June $16^{th}$, 2014. (a) S&P 500, (b) IPC, (c) Nikkei 225 and (d) XU 100.

The IPC index shows an increasing evolution of $C^*(t)$, suggesting a more uniform distribution of amplitudes at the beginning of 2014, Figure 7(b). On the contrary, the Nikkei 225 index presents a decreasing measure of complexity, indicating a period of higher complexity at the beginning of the sample period, Figure 7(c). Finally, the XU100 index presents alternate intervals of high and low complexity, displaying regularly large values in the last two months of the sample period. This higher randomness is also visible from the scaling exponent that displays relative lower values of $\bar{H}^*(t)$, Figure 7(d).

Overall, the functions $H^*(t)$ and $C^*(t)$ vary in opposite directions. For each market, the correlation between these two measures is negative with values $\rho_{S\&P} = -0.21$, $\rho_{IPC} = -0.23$, $\rho_{Nikkei} = -0.28$, $\rho_{XU} = -0.15$. We note that these correlations values are small, indicating a weak linear dependence between these variables. This is to be expected as the underlying measures are associated with rather different properties.

### *5.1. Intraday Analysis of Scaling Patterns*

We investigated the intraday patterns by separating the paths of $H^*(t)$ and $C^*(t)$ into daily windows. Taking for example the time series $H^*(t)$ for the S&P 500 index, Figure 6(a), we separated this time series into the $n = 105$ days that compose the data set, see Table 1. In Figure 8(a), we display these daily time series (one day on top of the other). The colour bar represents the value of $H^*(t)$. This graphical representation allows us to compare trading sessions and identify patterns at specific times of the day.

We estimated the statistical mean of $H^*(t)$ across the days, resulting in an average value for each time $t$ of the trading session. This average, denoted as $\langle H^*(t)\rangle_{days}$, describes the regular behaviour of $H^*(t)$ on a trading session, see Figure 8(b).

In order to validate that the observed dynamics of $\langle H^*(t)\rangle_{days}$ are statistically significant, we compared these dynamics with scaling exponents, $H^*_{BM}(t)$, obtained from several realizations of Brownian motions of length equal to the analysed financial time series, see Table 1. The time series of $H^*_{BM}(t)$ were fragmented into $n$ windows of equal length. The mean over these $n$ days is denoted as $\langle H^*_{BM}(t)\rangle_{days}$. The pink band reported in Figure 8(b) corresponds to the $5^{th}$ and $95^{th}$ percentile of the empirical distribution of $\langle H^*_{BM}(t)\rangle_{days}$ computed from 100 simulations.

We compared the values of $H^*(t)$ obtained for each financial index with the $\langle H^*_{BM}(t)\rangle_{days}$ band. At each time $t$ during the trading session, we estimated the relative fraction of $H^*(t)$ values that falls outside the pink band. In Figure 8(c), we report these results as a ratio of number of days outside the band divided by the total number of days. This ratio is labelled as likelihood. The colour bar of this figure represents the value of the average scaling exponent, i.e, $\langle H^*(t)\rangle_{days}$ (value plotted in Figure 8(b)). From this Figure, we observe that across the day there are periods of time with very high empirical probability of observing values of the scaling exponents significantly different from the corresponding values extracted from pure Brownian motion.

The mean of the complexity measure, $C^*(t)$, at each time $t$ of the trading session is shown in Figure 8(d). Equally as with the $H^*(t)$ exponent, the mean of $C^*(t)$ is computed across all $n$ days and it is denoted as $\langle C^*(t)\rangle_{days}$. The same daily analysis for the remaining three financial indices is reported in Figures 9, 10 and 11, respectively. Overall, from the intraday scaling and complexity measures we observed the following patterns:

- For each of the selected financial indices, the daily average $\langle H^*(t)\rangle_{days}$ displays an inverted U-shaped form that reflects a more chaotic behaviour at the beginning and at the end of the trading session. The opposite behaviour is observed for the complexity measure, $\langle C^*(t)\rangle_{days}$, which reveals a U-shaped form.

- The S&P 500 index displays the largest values of $H^*(t)$ (a stronger amplitude persistent behaviour) during the middle of the trading session. From Figure 8(a), we observe that most of the days present large values of $H^*(t)$ around midday. At this time, the average exponent $\langle H^*(t)\rangle_{days}$ reaches a value of 0.6, see Figure 8(b). These values are significantly different from what would be expected for Brownian motion with more than 80% of the observations outside the $5^{th}$ and $95^{th}$ percentile, Figure 8(c). Consistently, the complexity measure reaches its minimum at the same time of the trading session, see Figure 8(d).

- The Mexican stock exchange is characterized by some large scaling values at the middle of the day. However the most noticeable pattern is the large values just before the end of the trading session, see Figure 9(a). The mean of the windowed values reaches a maximum of 0.65, creating an upswing shape at this time, see Figure 9(b). This could be associated with an increase of trades in the last few minutes of the trading session that creates a more drastic change in the amplitudes. This pattern is only present in the Mexican stock exchange.

  From Figure 9(c), we observe that some minutes before the closing of the market, more than 90% of the local scaling indices fall outside the $5^{th}$ and $95^{th}$ percentile band for Brownian motion. The complexity measure also reflects a steep increase of disorder at the end of the trading session, see Figure 9(d).

- The Japanese and Turkish stock exchanges display two regions of large values for the scaling exponent. These regions are separated by the

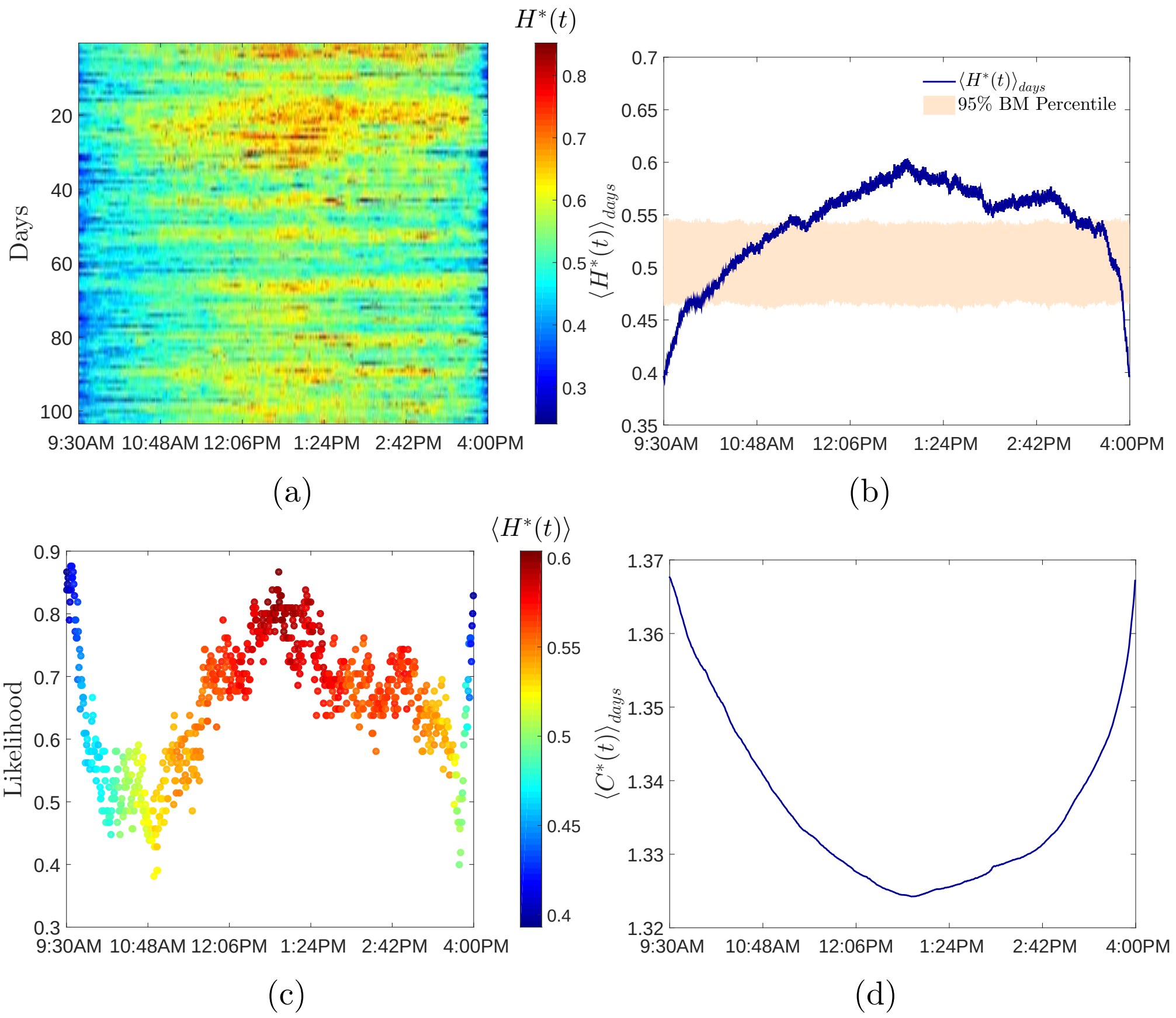

Figure 8: Intraday analysis for the S&P 500 Index.
(a) Intraday dynamics of the scaling exponent, $H^*(t)$, as a function of day and time. The colour bar indicates the value of the scaling exponent $H^*(t)$.
(b) Mean of $H^*(t)$ over the 105 days, denoted as $\langle H^*(t)\rangle_{days}$. The pink band corresponds to the $5^{th}$ and $95^{th}$ percentile of the distribution of $\langle H^*_{BM}(t)\rangle$ computed from 100 simulations.
(c) Likelihood of $H^*(t)$ to fall outside the $5^{th}$ and $95^{th}$ percentile band for Brownian motion (pink band of Figure (b)). The colour bar indicates $\langle H^*_{BM}(t)\rangle_{days}$, the value shown in Figure (b).
(d) Mean of the windowed complexity measure, denoted as $\langle C^*(t)\rangle_{days}$.

lunch break, see Figures 10(a) and 11(a) respectively. The mean of the scaling exponent reflects a quasi-double inverted U-shaped form that is associated with the opening and closing of the morning and afternoon sessions, Figures 10(b) and 11(b). It is worth noting that the two trading sessions do not display exactly the same profile. For the Japanese financial market, the inverted U-shaped form of the first trading session is slightly skewed to the right, in comparison with the more symmetric shape of the second trading session, see Figure 10(b).

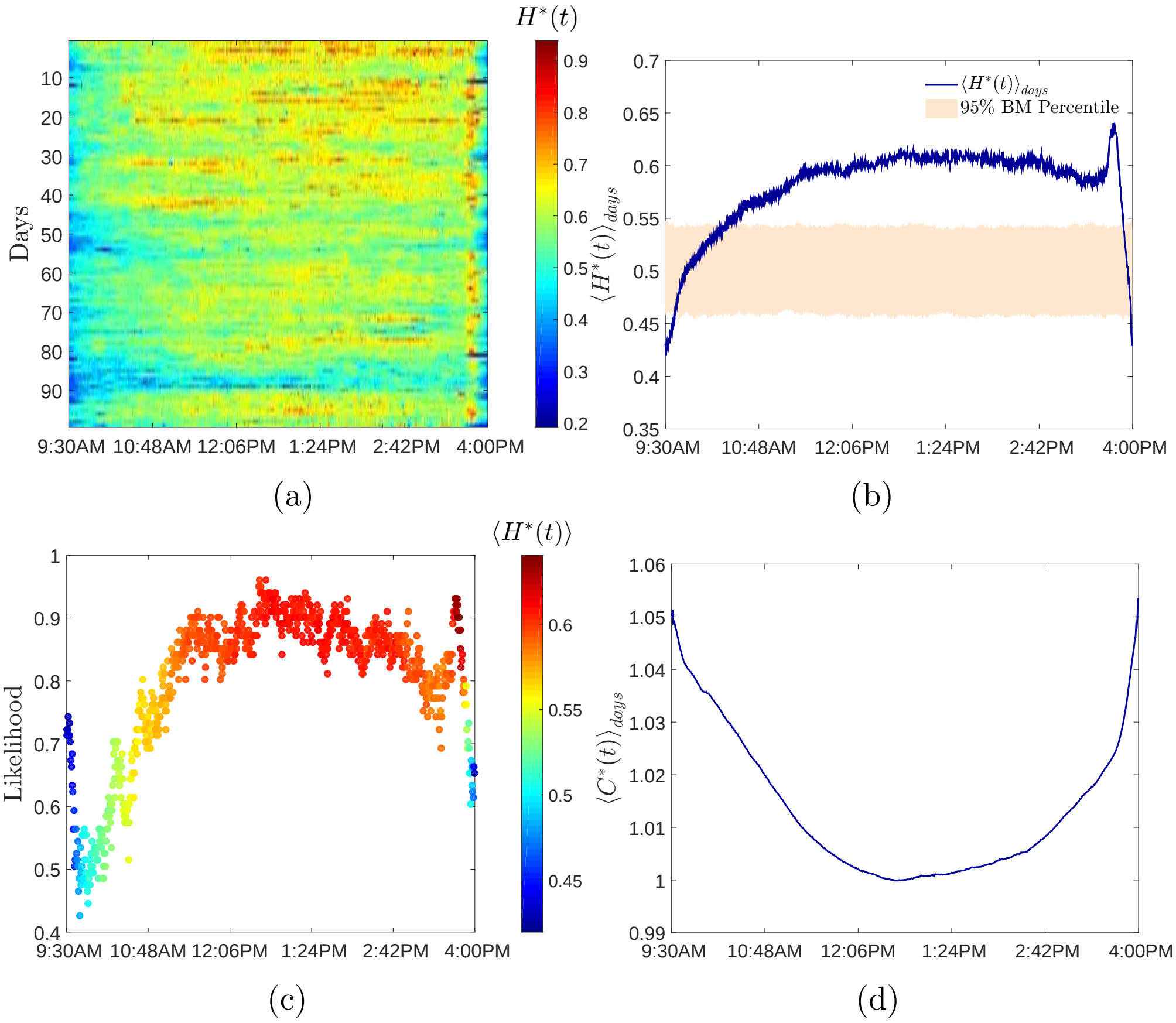

Figure 9: Intraday analysis for IPC. Caption for sub-figures (a), (b), (c) and (d) same as Figure 8.

- For the Turkish stock exchange, we observe that the first part of the trading session presents larger values of $H^*(t)$. More than 90% of the analysed days present local scaling exponents that surpass the value of 0.6, see Figure 11(c). The dominance of one IMF amplitude in the first trading session is also reflected in the lower values of the complexity measure, which reaches the lowest value when compared to the other

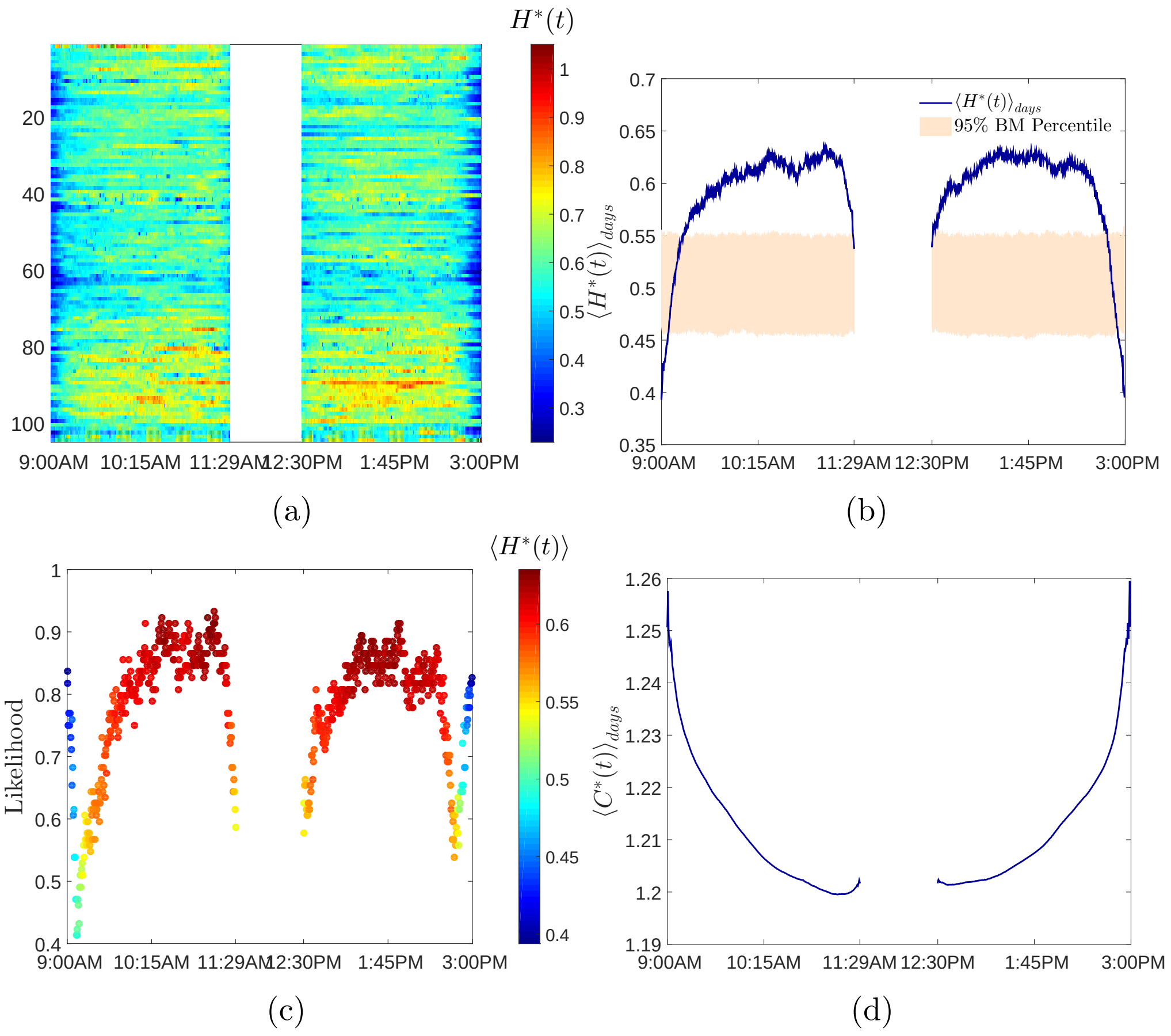

Figure 10: Intraday analysis for Nikkei 225 index. Caption for sub-figures (a), (b), (c) and (d) same as Figure 8. The white vertical band in each sub-figure corresponds to the lunch break in this stock exchange.

stock markets, see Figure 11(d).

Overall the intraday patterns of $H^*(t)$ and $C^*(t)$ confirm the well known fact that activity on financial markets is not constant throughout the day. The uncovered patterns corroborate the hectic buy and sell activity affecting different markets at the opening and closing of trading sessions [41, 42]. Smaller values of $H^*(t)$ (large values of $C^*(t)$) imply a non-persistent and rougher behaviour that is reflected in higher volatility. The exposed daily patterns of $\langle H^*(t)\rangle_{days}$ and $\langle C^*(t)\rangle_{days}$ are in agreement with the results that document the existence of a distinct U-shaped pattern in market activity and volatility over a trading day, i.e., volatility is higher at the opening and closing of trading and low in the middle of the day, see for example [43, 44, 45].

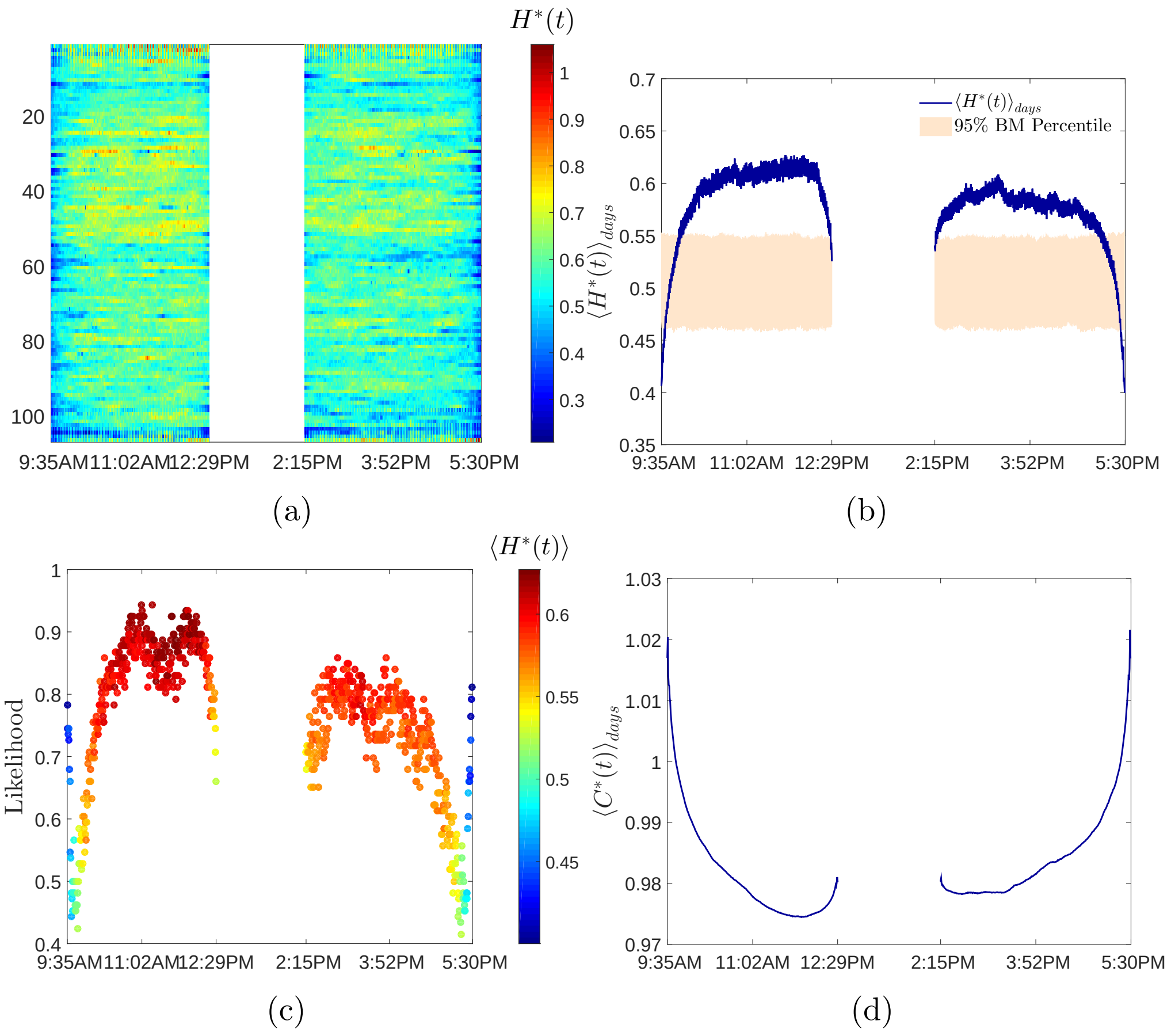

Figure 11: Intraday analysis for XU 100 index. Caption for sub-figures (a), (b), (c) and (d) same as Figure 8. The white vertical band in each sub-figure corresponds to the lunch break in this stock exchange.

By comparing the day-by-day complexity values across the four markets, we observe that S&P 500 index displays the largest values across all the trading session. Nikkei 225 is the second most complex, followed by IPC and lastly the XU100, which has the smaller complexity values. This is in agreement with the results reported for developed and emerging markets [11, 46, 22].

## 6. Conclusion

We studied the relative weight of the oscillating components present in intraday financial time series coupled with their characteristic time-scale. These components are extracted via the Hilbert-Huang transform. We have shown that the combination of EMD and its associated Hilbert spectral analysis offers a powerful tool to uncover the time-dependent scaling patterns of intraday financial data.

We proposed two new time-dependent measures: 1) an amplitude scaling exponent and 2) an entropy-like measure. Using these measures, we have been able to identify trends and intermittent behaviour in financial time series. Our measures are non-parametric and they do not assume any a priori stochastic process. The scaling exponent only assumes the existence of a power-law relation between the instantaneous amplitudes and the instantaneous periods that was empirically shown to be present.

We applied the scaling and the entropy-like measures to the study of four financial markets, two developed (US and Japan) and two emerging markets (Mexico and Turkey). We contrasted and compared the decomposition of their financial indices. With the use of intraday data, we recognized some patterns and identified periods of low and high complexity.

Compared to the other studied indices, the S&P 500 index results the most complex market. The intraday analysis reveals a distinctive anti-persistent behaviour at the opening and closing of the trading session, contrasting with the persistent behaviour at the middle of the session. Similar intraday results are obtained for the other stock indices. The variations observed in the scaling and complexity measures are well outside the $5^{th}$ and $95^{th}$ percentile expected for Brownian motion, suggesting strong deviations from this model that could be attributed to the presence of long range dependence or/and heavy tails.

With the proposed measures, we are able to describe the dynamics of financial time series whose regularity changes over time. Our results suggest that financial time series have dynamic scaling properties that could be attributed to the autocorrelation of the process, the presence of tails and the non-stationarity of the time series. Given the presence of this dynamic roughness, a time-varying scaling exponent could better reflect the scaling behaviour of financial data.

**Acknowledgement**

The authors wish to thank Bloomberg for providing the data. NN would like to acknowledge the financial support from Conacyt-Mexico. TDM wishes to thank the COST Action TD1210 for partially supporting this work. TA wishes to thank the Systemic Risk Centre at LSE.